\documentclass[aps,prl,amsmath,amssymb,twocolumn,superscriptaddress,showpacs,floatfix]{revtex4-1}
\usepackage{graphicx}
\usepackage{dcolumn}
\usepackage{bm}

\usepackage{xcolor}

\begin{document}
	
	\title{Phonon coupling between a nanomechanical resonator and a quantum fluid}
	
	\author{King Yan Fong}
	\affiliation{Department of Electrical Engineering, Yale University, New Haven, CT 06511, USA}
	\author{Dafei Jin}
	\affiliation{Center for Nanoscale Materials, Argonne National Laboratory, Lemont, IL 60439, USA}
	\author{Menno Poot}
	\affiliation{Department of Electrical Engineering, Yale University, New Haven, CT 06511, USA}
	\affiliation{Physik-Department, Technische Universit\"at M\"unchen, 85747 Garching, Germany}
	\author{Alexander Bruch}
	\affiliation{Department of Electrical Engineering, Yale University, New Haven, CT 06511, USA}
	\author{Hong X. Tang}
	\affiliation{Department of Electrical Engineering, Yale University, New Haven, CT 06511, USA}
	

	\begin{abstract}
		Owing to their extraordinary sensitivity to external forces, nanomechanical systems have become important tools for studying a variety of mesoscopic physical systems and realizing hybrid quantum systems. While nanomechanics has been widely applied in solid-state systems, its use in liquid is scantily studied. There it finds unique applications such as biosensing, rheological sensing, and studying fluid dynamics in unexplored regimes. Its use in quantum fluids offers new opportunities in studying fluids at low excitation levels all the way down to the quantum limit and in nano-metric scales reaching the fluid coherence length. Transduction and control of the low-loss excitations also facilitate long-life quantum information storage. In this work we demonstrate efficient coupling of a nanomechanical resonator to phonons in a bosonic quantum fluid -- superfluid $^4$He. By operating an ultra-high frequency nano-optomechanical microdisk resonator immersed in superfluid $^4$He, we show that the resonator dynamics is predominately determined by phonon-coupling to the superfluid. A high phonon exchange efficiency $>92\%$ and minimum excitation rate of 0.25 phonons per oscillations period are achieved. We further show that the nanomechanical resonator can strongly couple to superfluid cavity phonons with cooperativity up to 880. Our study opens up new opportunities in control and manipulation of superfluids in nano-scale and low-excitation level.
	\end{abstract}
	
	\maketitle
	
	
	The extreme sensitivity of nano-scale mechanical resonators to external forces allows them to couple to a wide range of physical systems such as photons at optical \cite{NatPhys_5_909_2009, Nature_459_550_2009} and microwave/radio-frequencies \cite{Nature_471_204_2011}, electron charge \cite{Nature_392_160_1998}, spin \cite{Nature_430_329_2004} and transport \cite{Science_325_1103_2009, Science_325_1107_2009}, defect centers in solids \cite{NatPhys_7_879_2011, Science_335_1603_2012}, superconducting qubits \cite{Nature_459_960_2009, Nature_464_697_2010}, SQUIDs \cite{NatPhys_4_785_2008}, and quantum dots \cite{NatNano_9_106_2014}. 
	This makes them a versatile tool for studying mesoscopic physics and realizing hybrid quantum systems \cite{PRL_105_220501_2010, PNAS_112_3866_2015}. In addition to confined resonant vibrations, traveling acoustic phonons also show potential for probing various systems and facilitate long-range communications between on-chip components \cite{NatNano_9_520_2014, Science_346_207_2014, NatCommun_5_5402_2014, NatPhotonics_10_346_2016}. 
	
	So far, while much effort has been made in applying nanomechanics in solid-state systems, very few works have addressed its use in the fluidic phase, where it enables, e.g,  studies of fluid dynamics in unexplored regimes \cite{PRL_92_235501_2004, PRL_98_254505_2007, RSI_84_025003_2013}, as well as biosensing and rheological sensing \cite{NL_15_6116_2015, Nature_446_1066_2007, NatCommun_4_1994_2013, APL_105_014103_2014, NatNano_10_810_2015, OE_25_821_2017}. Of particular interest is its use in superfluid $^4$He, which itself is a macroscopic quantum object with strong quantum correlation and nonlinearities. Since its monumental discovery dating back 80 years ago \cite{Nature_141_74_1938, Nature_142_643_1938}, superfluid $^4$He continues to attract a great deal of research interest till today. Sensing and understanding the elementary excitations in superfluid $^4$He are key to numerous fundamental studies such as detection of neutrinos \cite{PRL_58_2498_1987} and dark matters \cite{PRL_117_121302_2016}. 
	
	While macroscopic mechanical resonators have been used in the research of superfluid as transducers, e.g., vibrating wires \cite{PR_132_2373_1963}, 
	grids \cite{JLTP_158_462_2010}, 
	crystal oscillators \cite{PRB_79_054515_2009}, 
	and recently MEMS resonators \cite{JLTP_162_661_2011, JLTP_171_200_2012, JLTP_183_284_2016}, 
	their relatively large size and low oscillation frequency, however, limit their use to probing the average thermodynamic properties of the superfluid at macroscopic scales only. In this respect, nanomechanical systems offer new opportunities for studying superfluid dynamics in low-excitation level down to the quantum limit and at length scales comparable to, or even below, the superfluid coherence length \cite{SciRep_7_4876_2017}. Transduction and control of the low-loss excitations in the superfluid also have a great potential for long-life quantum information storage. Attempts of operating nanobeam resonators in liquid helium were made \cite{Nanotech_11_165_2000, SciRep_7_4876_2017} but the low quality factor $Q\sim 10$ even in the superfluidic phase \cite{SciRep_7_4876_2017} hinders their usage as high quality transducers. Recently, there has been a rising interest in utilizing superfluid $^4$He in optomechanical systems \cite{NJP_16_113020_2014, PRApp_7_044008_2017, NatPhys_12_788_2016, NatPhys_13_74_2017} 
	to take advantage of its low-loss property as a mechanical resonator. So far those systems, however, suffer from low optomechanical coupling rate which undermines the efficiency of optomechanical cooling and control. 
	\begin{figure*}[t]
		\centering
		\includegraphics[width=0.75\textwidth]{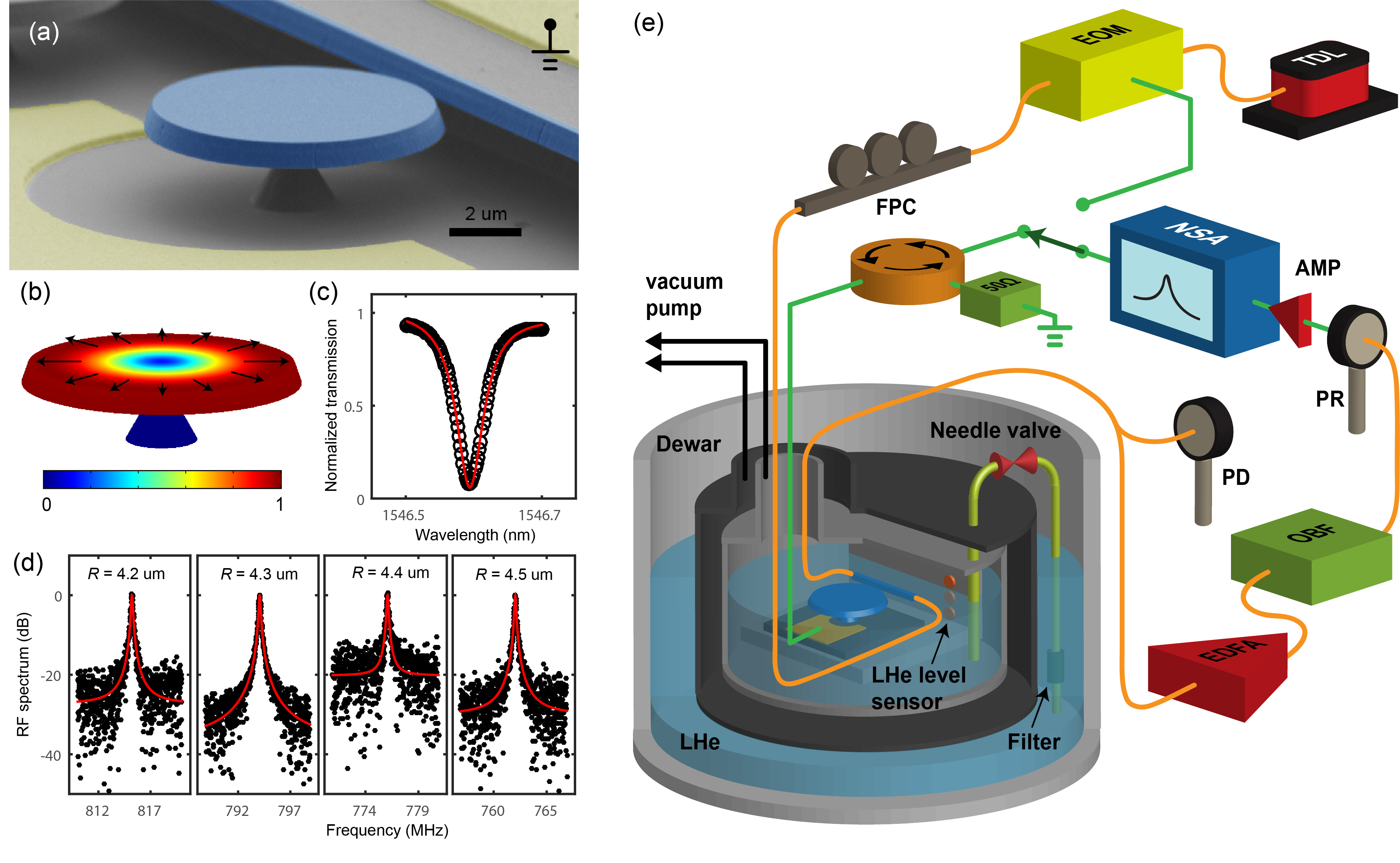}
		\caption{ \textbf{Device and measurement setup.} (a) SEM image of a fabricated device with false color. The microdisk and waveguide are shaded in blue and the electrodes are shaded in yellow. (b) Normalized displacement profile of the fundamental radial breathing mode computed using FEM. The arrows show the direction of the displacement. (c) Spectrum of a measured optical resonance showing an optical quality factor of 30,000. (d) Normalized RF response of the radial breathing mode of the microdisks measured in vacuum at room temperature. (e) Schematic of the measurement setup. TDL: Tunable Diode laser. EOM: Electro-optic phase modulator. FPC: Fiber polarization controller. PD: Photodetector. EDFA: Erbium-doped fiber amplifier. OBF: Optical bandpass filter. PR: High-speed photoreceiver. AMP: Electrical amplifier. NSA: Network/spectrum analyzer. }
		\label{fig1}
	\end{figure*}

	Here in this work we demonstrate efficient coupling of a nanomechanical resonator to phonons in a bosonic quantum fluid -- superfluid $^4$He. By operating an ultra-high frequency nano-optomechanical microdisk resonator in superfluid $^4$He, we show that the resonator dynamics are predominately determined by the phonon-coupling to the superfluid. High quality factors of $Q\sim 850$ in liquid are demonstrated. 
	Our device achieves high phonon exchange efficiency over 92\% and a low excitation level with a minimum excitation of 0.25 phonons per oscillations period. Such a performance is superior to any of the traditional superfluid transducer technologies. Moreover, strong coupling with cooperativities up to 880 between the nanomechanical resonator and superfluid cavity phonons are within reach. This provides new opportunities to control and manipulate superfluids at the nano-scale with low-level excitations. With the current technology of operation of nanomechanical systems in quantum regime \cite{Nature_475_359_2011, Nature_478_89_2011}, our system provides an ideal platform for studying coupled dynamics of macroscopic quantum objects. The low loss of the superfluid phonons also facilitates long-life quantum information storage and enables long-range quantum acoustic communication.

	


	The device under study is a nano-optomechanical microdisk resonator made of piezoelectric AlN \cite{PRA_90_051801_2014, APL_106_161108_2015}; a scanning-electron micrograph of a fabricated device is shown in Fig.~\ref{fig1}~(a). (See Methods for details on the device fabrication.) Its mechanical radial breathing mode is coupled to the low-loss optical whispering gallery mode, which can be used to efficiently transduce the mechanical displacement into optical signals. The normalized displacement profile of the fundamental radial breathing mode simulated by a finite-element method (FEM) is shown in Fig.~\ref{fig1}~(b). 
	Fig.~\ref{fig1}~(c) shows a typical measured spectrum of an optical resonance which has an optical quality factor of around 30,000. By applying a radio-frequency (RF) signal to the integrated electrode fabricated right next to the microdisk, the device can be driven piezoelectrically. A series of microdisks with top radii of $R=4.2$, 4.3, 4.4, 4.5 um is coupled to a single optical waveguide. Since the resonance frequency is inversely proportional to the disk radius, the individual disks can easily be identified in the frequency domain. This multiplexing approach allows measuring multiple disks in a single experiment. The measured spectra of their fundamental radial breathing modes are shown in Fig.~\ref{fig1}~(d).

	\begin{figure*}[t]
		\centering
		\includegraphics[width=0.8\textwidth]{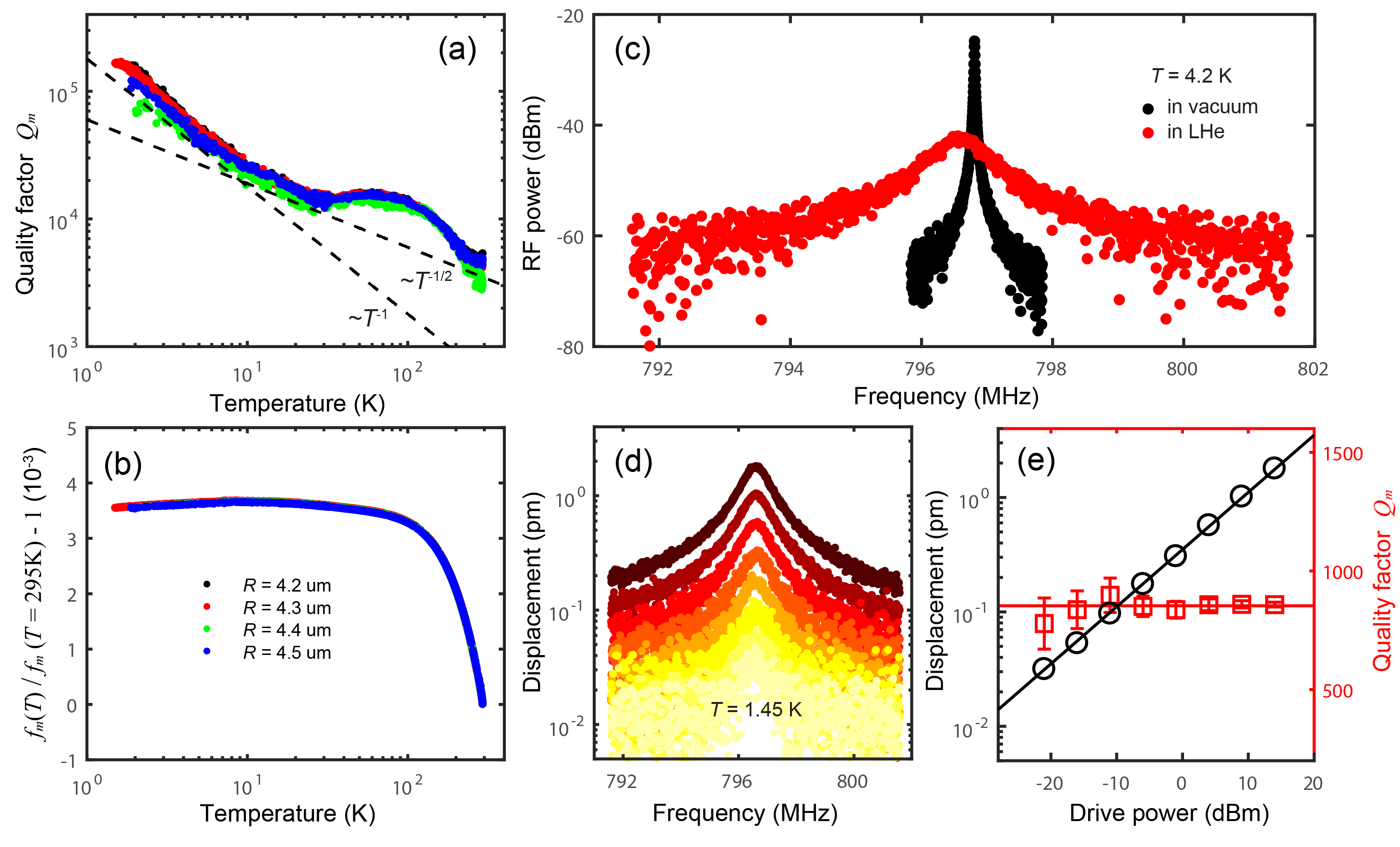}
		\caption{ \textbf{Device operation in vacuum and in liquid $^4$He.}(a) Mechanical quality factor $Q_m$ measured in vacuum plotted against temperature. The black dashed lines show the $T^{-1}$ and $T^{-1/2}$ dependence. (b) Relative frequency shift $f_m(T)/f_m(T=295\mathrm{K})-1$ measured in vacuum plotted against temperature. (c) Mechanical resonances measured in vacuum and in liquid $^4$He at $T=4.2$~K for the device with $R=4.3$~um. (d) Mechanical response measured at different driving power. 
			(e) The peak displacement and quality factor measured at different driving powers. The highest displacement attained is 1.78~pm. }
		\label{fig2}
	\end{figure*}

	The experimental setup is illustrated in Fig.~\ref{fig1}~(e); the details of the measurement scheme are explained in the Methods. The device performance was first measured in vacuum. At room temperature ($T=295$~K), the measured resonance frequencies are $f_m=815.2$~MHz, 794.0~MHz, 776.0~MHz, and 761.9~MHz, which follow the expected $1/R$ dependence (See Supp.~Info.). 
	The chip is cooled down to cryogenic temperatures, while continuously measuring the response of the four resonators. The extracted mechanical quality factor $Q_m$ and fractional shift of resonance frequency $\Delta f/f_m$ are plotted versus temperature in Fig.~\ref{fig2}~(a) and (b). As the temperature is lowered, the frequencies increase and reach a maximum at around 9~K before receding back slightly. The quality factors show a more dramatic increase from around 4,400 at room temperature to 170,000 at 1.45~K. Below 9~K a $T^{-1}$-dependence of $Q_m$ can be observed while above 9~K $Q_m$ follows $T^{-1/2}$ with a broad hump appearing at $T=$~30---200K. 
	
	Next, the device is immersed in liquid $^4$He. Now, both the resonance frequency and quality factor are reduced, as shown in Fig.~\ref{fig2}~(c). At $T=4.2$~K, $Q_m$ has dropped by almost a factor of 60, to about 1,000. On the other hand, $f_m$ only shows a minute down-shift of 0.03\%. 
	Fig.~\ref{fig2}~(d) and (e) plot the frequency response, the peak displacement and the quality factor at various driving powers measured at $T=1.45$~K. (Calibration of the actual displacement is discussed in the Methods and Supp.~Info..) 
	It shows that the displacement remains linear to the drive and that the quality factor stays constant up to the largest displacement of $A=1.78$~pm attained in the experiment. This corresponds to a peak velocity of $\omega A=8.9$~mm/s. 
	
\begin{figure*}[t]
		\centering
		\includegraphics[width=0.8\textwidth]{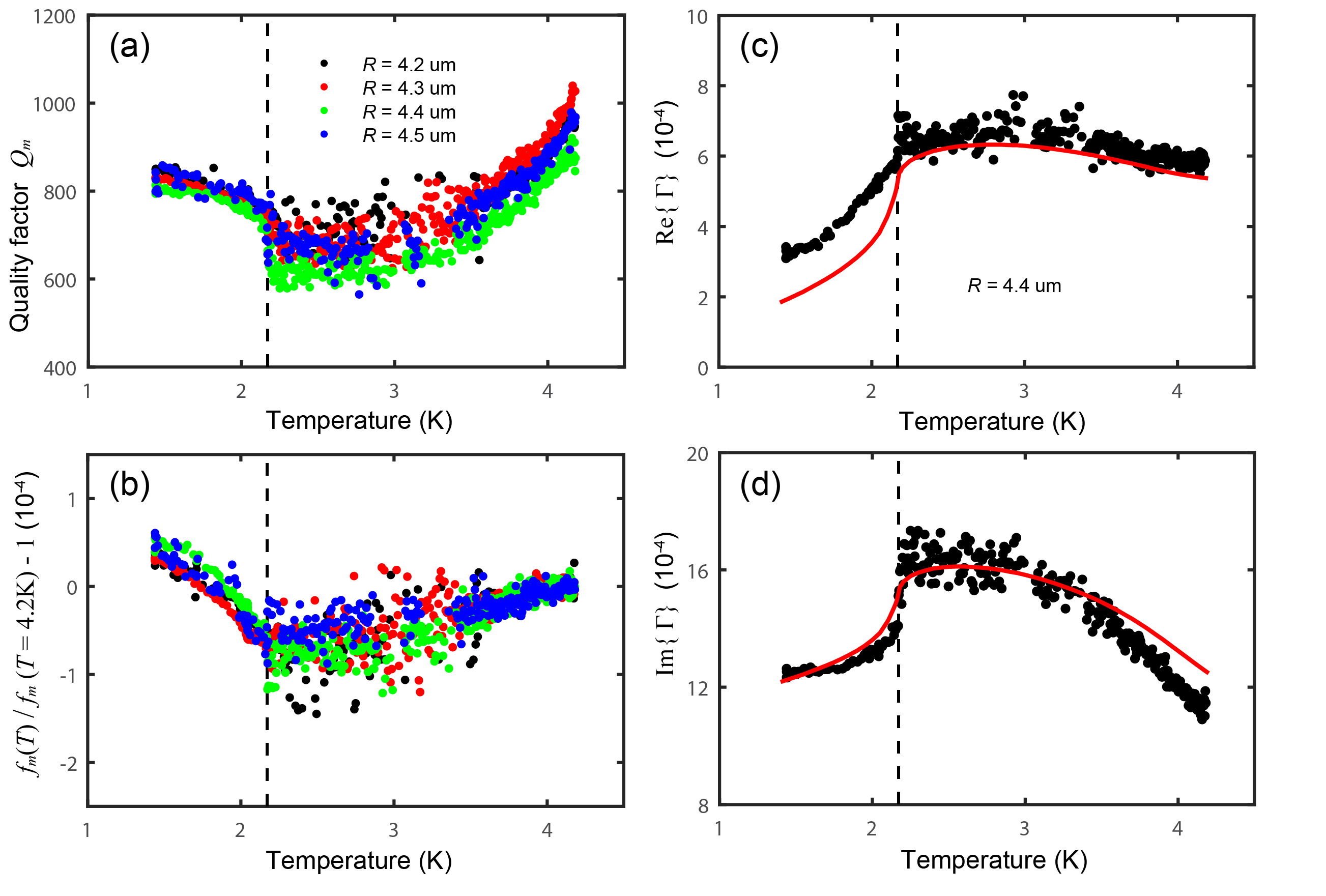}
		\caption{ \textbf{Measured hydrodynamic behavior compared with theory.} (a) Mechanical quality factor $Q_m$ and (b) relative frequency shift $f_m(T)/f_m(T=4.2\mathrm{K})-1$ measured in liquid $^4$He plotted against temperature for microdisks of various radius. The black dashed line indicates the $\lambda$-point $T=2.17$~K. (c) Real part $\mathfrak{Re}\{\Gamma\}$ and (d) imaginary part $\mathfrak{Im}\{\Gamma\}$ of the hydrodynamic function plotted versus temperature for $R=4.4~\mathrm{\mu m}$. The red solid lines represents the theoretical values based on the numerical model. }
		\label{fig3}
	\end{figure*}

	Subsequently, the temperature dependence of the quality factor and the resonance frequency were measured in LHe and are plotted in Fig.~\ref{fig3}~(a) and (b). In these measurements, an optical power of $-17$~dBm and RF power of $-15$~dBm were used. In general, the frequency and quality factor vary smoothly with temperature, but a clear kink in both $f_m$ and $Q_m$ is observed at $T=2.17$~K. This is the temperature of the $\lambda$-point, where the phase transition of normal fluid to superfluid $^4$He takes place. While the quality factor shows an increasing trend as the temperature drops below the $\lambda$-point, the rise slowly saturates resulting a quality factor at $T=1.45$~K even lower than that at $T=4.2$~K. This is counter-intuitive because at $T=1.45$~K over 90\% of the $^4$He is in the superfluid phase, so one would naively expect that the losses should be greatly reduced.

	To understand this unexpected effects and other behaviors of the device, we analyze the dynamics of the system using the Caldeira-Leggett model (CL model) \cite{PRL_46_211_1981}, which allows quantum and classical treatments of dissipation on equal footing. In the CL model, a mechanical resonator is described as a system coupled to a bath of harmonic oscillators. The Hamiltonian is given by
	\begin{align}
		&H_{sys} = H_{mech} + H_{bath} + H_{fluid} + H_{ex}
		\label{eq_H} \\
		&H_{mech} = \frac{p^2}{2m_e} +\frac{m_e\Omega_m^2 x^2}{2} 
		\label{eq_H_mech} \\
		&H_{bath} = \sum_k \left[ \frac{p_{b,k}^2}{2m_{b,k}} +\frac{m_{b,k}\omega_{b,k}^2}{2} \left(q_{b,k}-\frac{C_{b,k} x}{m_{b,k}\omega_{b,k}^2} \right)^2 \right]
		\label{eq_H_bath} \\
		&H_{fluid} = \sum_k \left[ \frac{p_{f,k}^2}{2m_{f,k}} +\frac{m_{f,k}\omega_{f,k}^2}{2} \left(q_{f,k}-\frac{C_{f,k} x}{m_{f,k}\omega_{f,k}^2} \right)^2 \right] 
		\label{eq_H_sf} \\
		&H_{ex} = -F_{ex} x ~.
	\end{align}
	$H_{mech}$ is the Hamiltonian for the mechanical resonator, where $p$ is the momentum conjugate to the displacement $x$, $m_e$ and $\Omega_m$ are the effective mass and resonance frequency of the resonator. $H_{fluid}$ is the Hamiltonian for the degrees of freedom provided by the fluid ($q_{f,k}$ and the corresponding momentum $p_{f,k}$). $H_{bath}$ is the Hamiltonian for the degrees of freedom outside the fluid ($q_{b,k}$, $p_{b,k}$). $H_{ex}$ is the Hamiltonian for the external drive. In Eqs.~(\ref{eq_H_bath}) and (\ref{eq_H_sf}), $C_{b,k}$ and $C_{f,k}$ represent the strength of the couplings between the resonator and the environments. 
	
	From the CL Hamiltonian, it can be derived using the Heisenberg picture that the mechanical resonator follows the equation of motion (see Supp.~Info. for details):
	\begin{align}
		m_e\ddot{x}+\frac{m_e\Omega_m}{Q_i}\dot{x}+m_e\Omega_m^2 x = F_{f} +F_{ex} +\xi ~.
		\label{eq_x}
	\end{align}
	$Q_i$ is the intrinsic quality factor of the resonator in absence of the superfluid. It originates from $H_{bath}$ that represents the dissipation to the degrees of freedom outside the superfluid. $F_{ex}$ is the external driving force and $\xi$ is the thermal fluctuation force. $F_{f}$ represents the force acting on the resonator caused by the coupling to the superfluid. In the frequency domain it can be written as
	\begin{align}
		F_f[\omega] =x[\omega] m_e \omega^2 \Gamma[\omega]
		\label{eq_F_f}
	\end{align}
	where
	\begin{align}
		&\Gamma[\omega] = i\frac{J_f[\omega]/m_e}{\omega^2} - \frac{2}{\pi} \int_0^\infty d\omega^\prime \frac{J_f[\omega^\prime]/m_e}{\omega^\prime (\omega^{\prime2}-\omega^2)}
		\label{eq_Gamma_J} \\
		&J_f[\omega] = \frac{\pi}{2}\sum_k^N\frac{C_{f,k}^2}{m_{f,k} \omega_{f,k}} \delta(\omega-\omega_{f,k}) ~.
		\label{eq_J}
	\end{align}
	$\Gamma[\omega]$, as defined by Eq.~(\ref{eq_F_f}), is the dimensionless ``hydrodynamic function'' commonly used in hydrodynamic models for mechanical resonators in fluid \cite{PRL_92_235501_2004, NL_15_6116_2015, JAP_84_64_1998, PhysFluids_21_013104_2009}. (The definition adopted here differs from those in Refs.~\cite{JAP_84_64_1998, PRL_92_235501_2004} by a geometrical factor.) $J_f[\omega]$, as defined by Eq.~(\ref{eq_J}), is the spectral density of the bath modes of the superfluid determined by the CL model parameters. These two quantities from models of different perspectives are directly related through Eq.~(\ref{eq_Gamma_J}).
	
	Combining Eqs.~(\ref{eq_x}) and (\ref{eq_F_f}), it can be shown that the real part of the hydrodynamic function $\mathfrak{Re}\{\Gamma\}$ represents the mass-loading effect that modifies the resonance frequency $\Omega_f=\Omega_m/\sqrt{1+\mathfrak{Re}\{\Gamma\}}$ and the imaginary part $\mathfrak{Im}\{\Gamma\}$ represents the fluidic damping that modifies the quality factor $Q_f^{-1} \approx Q_i^{-1} + \mathfrak{Im}\{\Gamma\}$. Experimentally, the measured frequency shift (relative to that in vacuum) and quality factor can be used to extract $\Gamma[\omega]$ and thus $J_f[\omega]$. Fig.~\ref{fig3}~(c) and (d) plot the extracted  $\mathfrak{Re}\{\Gamma\}$ and $\mathfrak{Im}\{\Gamma\}$ for $R=4.4~\mathrm{\mu m}$. 
	
	Numerically, it is possible to compute the fluidic force from a hydrodynamic model to obtain $\Gamma[\omega]$ and $J_f[\omega]$. Here, we develop a hydrodynamic model based on Refs.~\cite{JEngMath_3_29_1969, JAP_84_64_1998, PhysFluids_21_013104_2009}. 
	Details of the model can be found in the Supp.~Info.. Consider the following two coupled processes: (I) the nanomechanical resonator oscillates, which actuates the flow of the surrounding superfluid, and (II) the superfluid evolves and in turn exerts a force on the resonator through momentum exchange. A self-consistent solution is obtained by combining these two processes.
	
	To analyze the actuation of the superfluid by the resonator, we start with the phenomenological Landau-Khalatnikov two-fluid model \cite{Book_Khalatnikov_1965}. 
	The linearized equations of motion are given by:
	\begin{align}
		&\frac{\partial\delta\rho}{\partial t} 
		+\nabla\cdot\vec{J} =0
		\label{eq_mass_linear} \\
		&\frac{\partial\vec{J}}{\partial t}  
		+\nabla \delta P -\eta\nabla^2\vec{u}_n -\left(\frac{\eta}{3}+\zeta_2\right) \nabla(\nabla\cdot\vec{u}_n) \nonumber \\
		&~~~~~~~~~~~~ -\rho_{s0}\zeta_1 \nabla[\nabla\cdot(\vec{u}_s-\vec{u}_n)] =0 \label{eq_momentum_linear} \\
		&\rho_0 \frac{\partial \delta s}{\partial t} +s_0 \frac{\partial\delta\rho}{\partial t} +\rho_0 s_0 \nabla\cdot\vec{u}_n -\frac{\kappa\nabla^2 \delta T}{T_0} = 0
		\label{eq_entropy_linear} \\
		&\frac{\partial\vec{u}_s}{\partial t}
		+\frac{1}{\rho_0}\nabla \delta P -s_0 \nabla \delta T -\zeta_4 \nabla(\nabla\cdot\vec{u}_n) \nonumber \\
		&~~~~~~~~~~~~ -\rho_{s0}\zeta_3 \nabla[\nabla\cdot(\vec{u}_s-\vec{u}_n)] =0 
		\label{eq_superfluid_linear}
	\end{align}
	where $\vec{u}_n$ ($\vec{u}_s$) and $\rho_n$ ($\rho_s$) are the velocity field and density of the normal fluid (superfluid) component, $\rho=\rho_n+\rho_s$ is the total density, $\vec{J}= \rho_n\vec{u}_n +\rho_s\vec{u}_s$ is the total momentum density, $P$, $T$ and $s$ are the pressure, temperature and specific entropy, $\eta$ is the dynamic viscosity, $\zeta_1$, $\zeta_2$, $\zeta_3$, $\zeta_4$ are the second viscosity coefficients,  and $\kappa$ is the thermal conductivity. For the quantities $z\in\{P,T,s,\rho\}$, $z_0$ denotes the values at thermal equilibrium and $\delta z$ denotes the small time-varying component induced by the resonator's motion $x(t)$. Eqs.~(\ref{eq_mass_linear})--(\ref{eq_superfluid_linear}) describe respectively the flow of mass, momentum, entropy and the superfluid component. Together with the thermodynamic relations between $P$, $T$, $s$, and $\rho$, 
	%
	%
	%
	%
	they form a complete set of partial differential equations (PDEs). The effect of the vibrations of the nanomechnical resonator is included through oscillatory boundary conditions.
	
\begin{figure*}[t]
		\centering
		\includegraphics[width=0.8\textwidth]{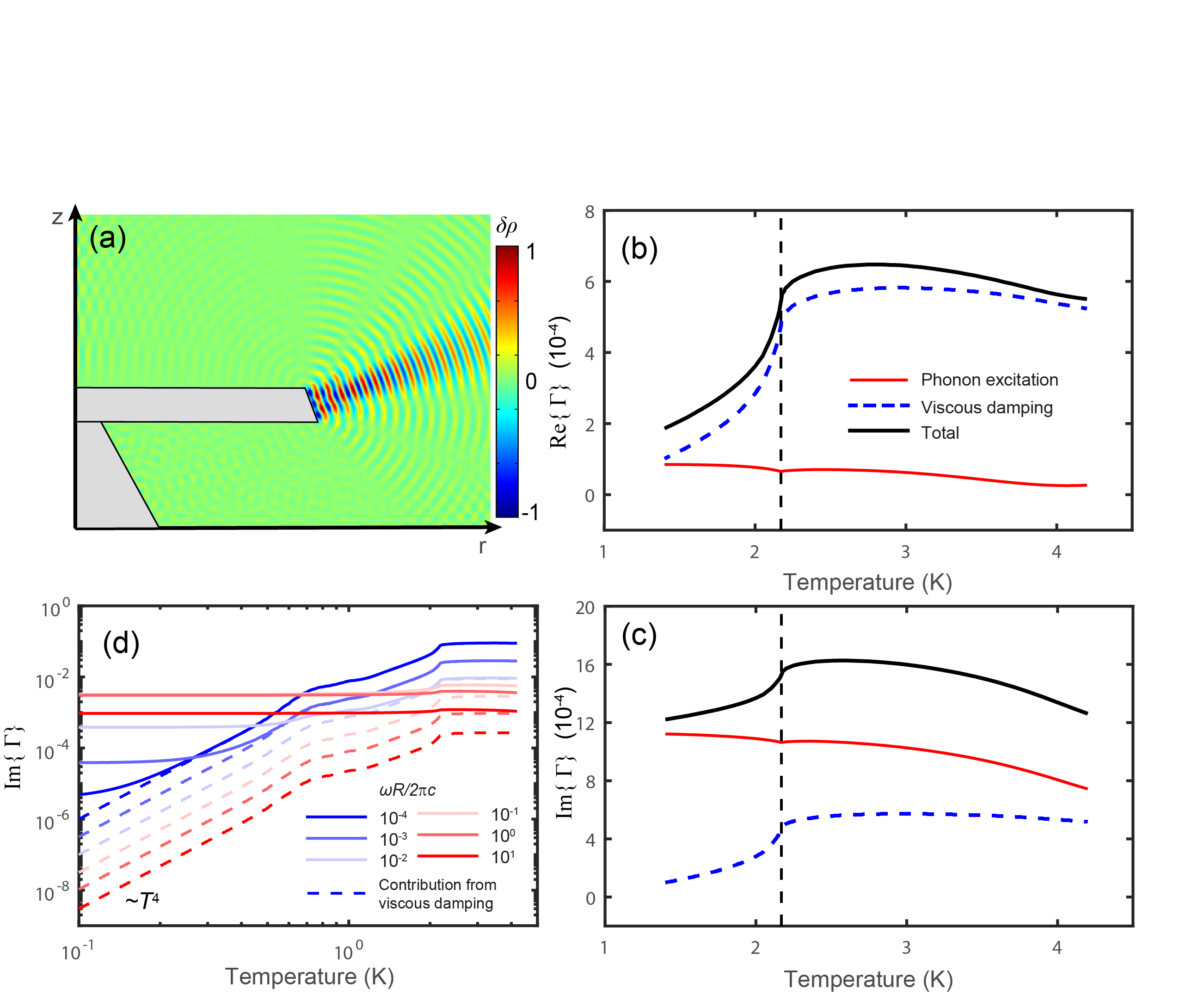}
		\caption{ \textbf{Predictions of the theoretical model.} (a) Color-plot of the simulated liquid $^4$He phonon wave in the vicinity of the nanomechanical resonator in cylindrical coordinates. The color scale represents the density field of the $^4$He. The gray region is the microdisk resonator, which has a thickness of $0.65~\mathrm{\mu m}$ and top-radius of $4.4~\mathrm{\mu m}$.  (b) Real part $\mathfrak{Re}\{\Gamma\}$ and (c) imaginary part $\mathfrak{Im}\{\Gamma\}$ of the hydrodynamic function plotted against temperature. The black dashed line indicates the $\lambda$-point. (d) $\mathfrak{Im}\{\Gamma\}$ plotted against temperature for various $\omega R/c$ ratios. The contribution from the shear viscous damping is plotted in dashed lines. }
		\label{fig4}
	\end{figure*}

	Because of the very high kinetic Reynold's number $\mathrm{Re}=\rho_n\omega R^2/\eta \sim 10^6$, the PDE problem can be treated first using an inviscid model (i.e. with $\eta=0$) to account for the acoustic phonon excitation, and then using the boundary-layer method to account for the shear viscous effects. The dynamic Reynold's number $\mathrm{Re}_D=\rho_n\omega A R/\eta < 1$ remains small so inertial effects such as non-laminar flow can be neglected. This is supported by the observed independence of the quality factor on the motion amplitude in Fig.~\ref{fig2}~(e). We solve the resulting PDE problem using FEM. A calculated normalized density field $\delta\rho$ of the phonon waves generated around the resonator is shown in Fig.~\ref{fig4}~(a). 
	
	For the back-action from the superfluid to the resonator, 
	a detailed analysis (See Supp.~Info.) shows that the fluidic force 
	$F_f=F_p+F_v$ can be separated into contributions from the phonon pressure force $F_p$ and the shear viscous force $F_v$. These are given by
	\begin{align}
		F_p &= -c^2\oint_S dA \left(\varphi_\perp^{*} \delta\rho \right)
		\label{eq_Gamma_p} \\
		F_v &= c^2 \frac{(1+i)\delta_e}{\sqrt{2}} \oint_S dA  \left[ \varphi_\parallel^{*} \left(\varphi_\parallel\rho_n k_{ph}^2 x -\frac{\rho_{n0}}{\rho_0} \nabla_\parallel\delta\rho\right) \right] ~,
		\label{eq_Gamma_v}
	\end{align}
	where 
	$c$ is the speed of first sound, $k_{ph}=\omega/c$ is the phonon wave-vector, $\varphi_\perp$ and $\varphi_\parallel$ are normal and tangential components of the resonator eigenmode displacement at the surface, and $\delta_e$ is the effective boundary length. 
	The integrals are carried out over the resonator's surface. Note that $\delta\rho$ and $\varphi_\perp$ are generally not in phase. $F_p$ originates from the inviscid model and the work done by $F_p$ represents the energy exchange between the resonator and the superfluid phonons. $F_v$ is a dissipative force which leads to energy loss of the system. 

	The results of the two coupled processes can be combined to obtain a self-consistent solution for $\Gamma[\omega]$. Fig.~\ref{fig3}~(c) and (d) plot the computed hydrodynamic functions based on the theoretical model, which agrees excellently with the experimental results. The model correctly reproduces the temperature dependence of both $\mathfrak{Re}\{\Gamma\}$ and $\mathfrak{Im}\{\Gamma\}$ and obtains values in the right order of magnitude. 
	In the calculation, the material parameters used are the experimentally measured values \cite{JPhysChemRefData_27_1217_1998}. The effective boundary length $\delta_e=K\delta_B$ is found to be $K=3.7$ times the Stoke boundary length $\delta_B=\sqrt{\eta/\rho_{n0}\omega}$.  ($\delta_B\sim$ 2 -- 5~nm throughout the temperature range.) We attribute this increase in effective boundary length to the surface roughness of the resonator, which is expected to cause additional fluid trapping at the surface and hence stronger viscous effects \cite{AnalChem_65_2910_1993}. 
	A further refinement of the model will be the inclusion of the side-coupling waveguide and the bottom electrodes.
	
	To gain more physical insight, we plot the hydrodynamic functions contributed from phonon excitation $\Gamma_p$ and from viscous damping $\Gamma_v$ (calculated from Eq.~(\ref{eq_Gamma_p}) and (\ref{eq_Gamma_v}) respectively) in Fig.~\ref{fig4}~(b) and (c). 
	One can see that at low temperature the viscous effect is reduced significantly while the effect of phonon excitation stays strong, resulting in a higher dissipation ($\mathfrak{Im}\{\Gamma\}\approx Q_m^{-1}$) at 1.45~K than at 4.2~K. At $T=1.45$~K, the effect of phonon excitation accounts for over $\mathfrak{Im}\{\Gamma_p\} /\mathfrak{Im}\{\Gamma\} =92\%$ of the overall dissipation of the resonator. For the device with $R=4.2$~um, $f_m=817.7$~MHz and $Q_m$ of $850$, the average number of thermal phonon is $\bar{n}_{th}=[\exp(hf_m/k_B T)-1]^{-1} =37$. The dissipation rate corresponds to an average exchange of 0.25 phonons per period of oscillation between the nanomechanical resonator and the superfluid. This sets the minimum operation level of the device as a phonon transducer. The high efficiency $>92\%$ and low excitation rate of few phonons per oscillation period is superior to any of the traditional transducer technologies \cite{PRB_79_054515_2009, JLTP_158_462_2010}. This technique also enables pure phonon generation that dominates over viscous dissipation. 

	The efficient phonon excitation in our system can be attributed to the large $\omega R/c$ ratio, which is analogous to the Mach number in non-oscillatory flows. 
	This is illustrated in Fig.~\ref{fig4}~(d), which plots $\mathfrak{Im}\{\Gamma\}$ 
	versus temperature for different $\omega R/c$. Also shown is the viscous contribution. In all cases, the latter drops quickly at low temperature as $\sim T^4$. 
	As $\omega R/c$ increases, the viscous effect is reduced while the phonon excitation effect rises, eventually the latter dominates the dissipation in the whole temperatures range. This behavior could provide an explanation to the saturation of resonator dissipation at low temperature observed in previous study \cite{JLTP_171_200_2012}.
	
\begin{figure}[t]
		\centering
		\includegraphics[width=0.45\textwidth]{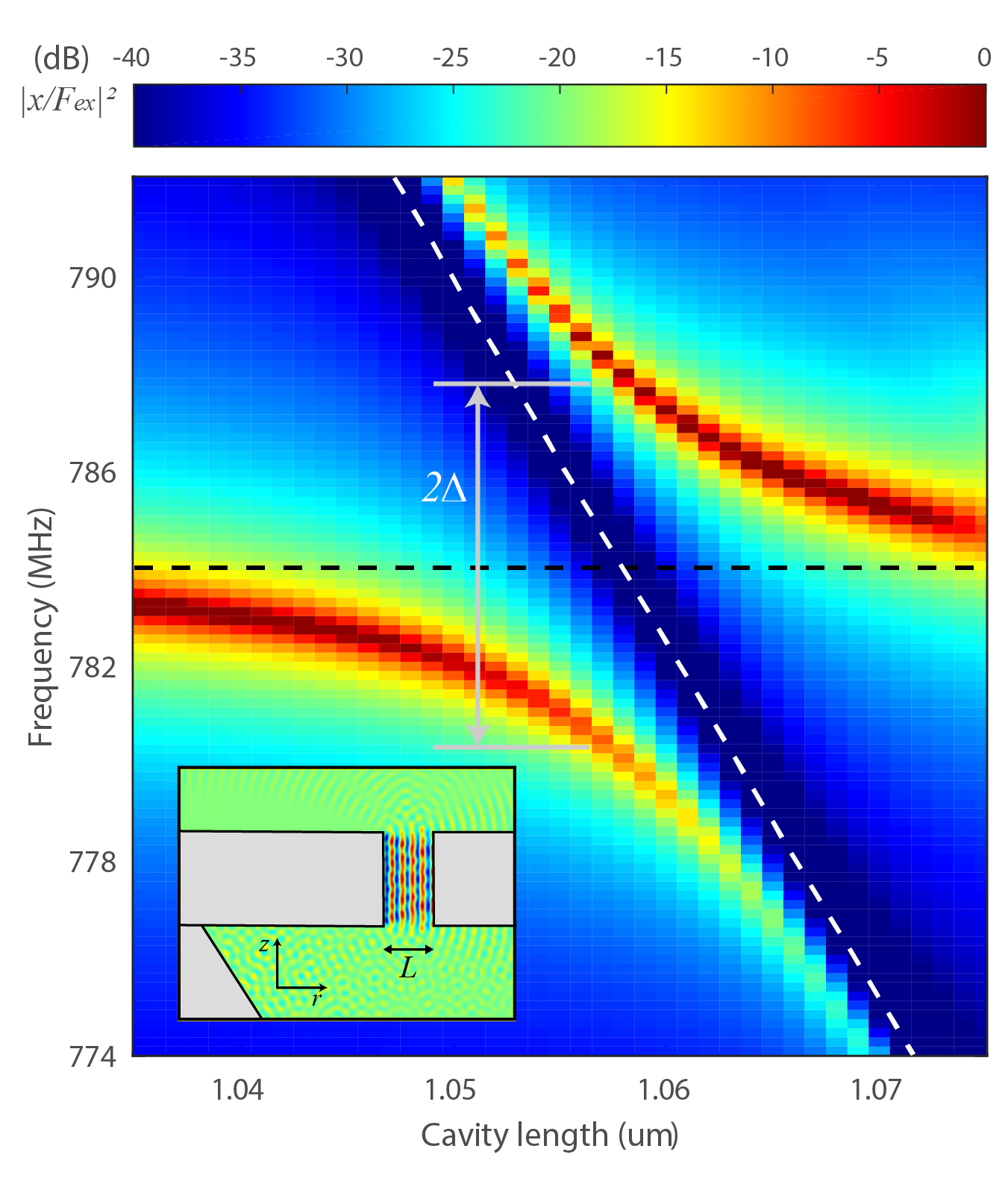}
		\caption{ \textbf{Strong coupling between nanomechanical resonator and superfluid phonon} Color plot of the device response $|x[\omega]/F_{ex}[\omega]|^2$ as a function of frequency and cavity length $L$. The black and white dashed lines represent the uncoupled resonance frequencies of the nanomechanical resonator and superfluid $^4$He cavity phonon. Inset shows the simulated density field of the phonon in superfluid $^4$He. The microdisk has a thickness of $2~\mathrm{\mu m}$ and radius of $4.4~\mathrm{\mu m}$. }
		\label{fig5}
	\end{figure}

	The efficient phonon excitation in $^4$He also suggests that it is possible to realize strong coupling between the nanomechanical phonons and \emph{confined} superfluidic phonons. To illustrate this, we simulate a system consisting of a microdisk and an outer ring structure that prevents the phonon from radiating energy away. The gap forms an open phonon cavity with cavity length $L$ (see inset of Fig.~\ref{fig5}). Now since only the single cavity phonon mode is of concern, we can rewrite the Hamiltonian Eq.~(\ref{eq_H}) as \cite{footnote_1}
	\begin{align}
		H =& \hbar\Omega_m a^\dagger a + \hbar\Omega_{cav} b^\dagger b \nonumber \\ 
		&+\hbar g(a^\dagger + a)(b^\dagger + b) + H_{bath}^{\prime} +H_{ex}
		\label{eq_H_cav}
	\end{align}
	where $a$ ($a^\dagger$) and $b$ ($b^\dagger$) are the creation (annihilation) operators for the mechanical resonator and the particular cavity phonon mode. Other degrees of freedom in superfluid are lumped into $H_{bath}^{\prime}$. When $\Omega_{cav}$ is close to $\Omega_m$, which can be tuned by adjusting the cavity length $L$, the Hamiltonian in Eq.~(\ref{eq_H_cav}) can be diagonalized with eigenmodes that are linear combinations of $a$ and $b$. It means that the resonator mode and the cavity phonon become hybridized.
	Fig.~\ref{fig5} plots the simulated response of the mechanical resonator $|x[\omega]/F_{ex}[\omega]|^2$ for different $L$ at $T=1.4$~K. (Additional numerical results and analysis can be found in Supp.~Info..) For large frequency differences, the response is close to that of the uncoupled resonators (black dashed line) which has a linewidth of 
	$w_{\mathrm{mech}} = f_m/Q_m = 0.135 \mathrm{~MHz}$. Here the quality factor ($Q_m=5800$) is higher than those in the previous device design mainly because of the the presence of the outer ring structure that modifies the radiation of the phonon modes. 
	When the detuning is decreased by changing $L$, a response appears near the cavity phonon mode (white dashed line) with a linewidth of $w_{\mathrm{cav}} =  0.522 \mathrm{~MHz}$. In the color-plot a clear anti-crossing is visible, indicating  strong coupling between the resonator motion and superfluidic phonon. 
	Our simulations thus show that, even though the phonon is not confined at every side, the phonon exchange between the mechanical resonator and the superfluid is sufficiently strong to prevail that loss. From the frequency splitting $2\Delta = 7.87 \mathrm{~MHz}$, a single-phonon cooperativity $C=4\Delta^2/w_{\mathrm{mech}} w_{\mathrm{cav}} = 880$ is obtained. It is anticipated that an optimized design of the phonon cavity can further minimize the phonon radiation loss and enhance the coupling.

	In conclusion, in this work we experimentally study the phonon-coupling between nanomechanical resonators and superfluid $^4$He. We show that the dynamics of an ultra-high frequency nano-optomechanical microdisk resonator immersed in superfluid $^4$He is predominantly influenced by the phonon coupling to superfluid. We further show that strong coupling between the nanomechanical resonator and superfluid cavity phonons can be achieved with cooperativity up to 880. Our study opens up new opportunities in control and manipulation of superfluid in nano-scale and low-excitation level in integrated platform. It also provides an ideal platform for studying coupled dynamics of macroscopic quantum objects. While in this study the system is operated in linear regime, nonlinear regime where mutual friction between the two fluid components become significant will be an interesting topic for further investigation. 

	\section*{Author contributions}
	K.Y.F. performed the experiment and analyzed the data. K.Y.F. and D.F.J. developed the numerical model. M.P. and A.B. assisted in the measurement. H.X.T. supervised the project. All authors discussed the results and contributed to the writing of the manuscript.
	
	\section*{Competing financial interests}
	The authors declare no competing financial interests.
	
	\section*{Acknowledgements}
	H.X.T. acknowledges support from a Packard Fellowship in Science and Engineering and a career award from National Science Foundation. This work was funded by the DARPA/MTO ORCHID program through a grant from the Air Force Office of Scientific Research (AFOSR) and a STIR grant from Army Research Office (ARO). M.P. acknowledges support of the Technische Universität München – Institute for Advanced Study, funded by the German Excellence Initiative (and the European Union Seventh Framework Programme under grant agreement n° 291763).
	
	\section*{Methods}
	
	\subsection*{Device fabrication}
	
	The fabrication process started with a silicon wafer with 2~um thermal oxide, 650~nm magnetron-sputtered aluminum nitride, and 100~nm PECVD silicon dioxide. E-beam lithography was performed to define the device pattern and reactive-ion etching (RIE) was used to dry etch the top PECVD oxide and 400~nm of the aluminum nitride layer. The Cl$_2$-based RIE of AlN caused the tapered shape of the microdisk with a side-wall angle of $20^\circ$. A second e-beam lithography step was then performed to pattern the releasing windows where the remaining aluminum nitride layer was etched completely by RIE. Next, the bottom silicon oxide was etched in a buffered HF solution to partially suspend the mechanical structure. This etching was timed so that the 2~um silicon dioxide was removed, exposing the silicon substrate while the microdisk was not yet fully undercut. This ensures that the pedestal was sturdy enough to survive in the subsequent lithography and lift-off processes for patterning the integrated electrodes next to the resonator. Finally the device was again etched in a buffered HF solution to now fully release the microdisk structure followed by drying in a critical point dyer.
	
	\subsection*{Measurement scheme}
	
	Refer to Fig.~\ref{fig1}~(e) in main text. Inside a LHe cryostat equipped with a 1K pot, the device was aligned to an optical fiber array and a microwave probe for simultaneous optical and electrical access. Liquid $^4$He was introduced into the sample chamber through a tube connecting the chamber and the outside dewar. A filter with pore size of 0.5~um was use at the end of the tube to prevent contamination. The inflow of liquid $^4$He was controlled by a needle valve in the middle of the tube and by the chamber pressure. The LHe level was monitored by a series of sensors. When measuring the device immersed in liquid $^4$He, the LHe was kept at its saturation pressure. Outside the cryostat, laser light from a tunable diode laser was sent to the device through an electro-optic phase modulator and a fiber polarization controller. The phase modulator was used to calibrate the actual displacement of the resonator \cite{OE_18_23236_2010}. The optical signal coming from the device was amplified by an Erbium-doped fiber amplifier and collected by a high speed photoreceiver. Electrical driving signals from a network/spectrum analyzer were sent to the device through an RF circulator, which is connected to a $50~\mathrm{\Omega}$ load to absorb the reflected RF power.
	

\begin{thebibliography}{53}
\bibitem{footnote_1} It can be derived by substituting $a = x\sqrt{m_e\Omega_m/2\hbar} + ip/\sqrt{2\hbar m_e\Omega_m}$, $b = q_f\sqrt{m_f\omega_f/2\hbar} + ip_f/\sqrt{2\hbar m_f\omega_f}$ and $g=C_k/2\sqrt{m_e\Omega_m m_f\omega_f}$. 
\end{thebibliography}

\end{document}